\documentclass[10pt,conference]{IEEEtran}
\usepackage{multirow}

\usepackage{cite}
\usepackage{amsmath,amssymb,amsfonts}
\usepackage{algorithmic}
\usepackage{graphicx}
\usepackage{booktabs}
\usepackage{textcomp}
\usepackage{tabularx}
\usepackage{xcolor}
\usepackage[linesnumbered,ruled,vlined]{algorithm2e}
\usepackage[normalem]{ulem}

\usepackage{tikz}
\usetikzlibrary{shapes,arrows.meta,positioning,calc}
\usetikzlibrary{arrows.meta}

\tikzset{
  cloudnode/.style={
    cloud, draw=yellow!50!black, fill=yellow!10,
    text centered, text width=1.5cm,
    minimum height=0.4cm, aspect=1.9, font=\scriptsize
  },
  component/.style={
    rectangle, rounded corners=2pt,
    minimum width=2.2cm, text width=2.1cm,
    minimum height=0.8cm,
    text centered, draw=black, fill=blue!10, font=\scriptsize
  },
  tool/.style={
    rectangle, minimum width=2.2cm, minimum height=0.6cm,
    text centered, draw=black, fill=gray!20, font=\scriptsize
  },
  storage/.style={
    cylinder, shape border rotate=90,
    minimum height=0.4cm, minimum width=0.5cm,
    text centered, draw=black, fill=orange!20, font=\scriptsize
  },
  trigger/.style={
    diamond, aspect=1.4, draw=black, fill=red!20,
    minimum width=1.3cm, text centered, text width=1.6cm, font=\scriptsize
  },
  arrow/.style={->, thick, >=stealth},
  act/.style={
    font=\scriptsize\itshape,
    text=green!40!black,
    fill=green!20,
    draw=green!40!black,
    inner sep=2pt,
    midway, sloped
  }
}

\begin{document}

\title{Praxium: Diagnosing Cloud Anomalies with AI-based Telemetry and Dependency Analysis}

\author{\IEEEauthorblockN{Rohan Kumar, Jason Li, Zongshun Zhang, Syed Mohammad Qasim, Gianluca Stringhini, Ayse K. Coskun}
\IEEEauthorblockA{
\textit{Boston University}}
}


\maketitle

\begin{abstract}
As the modern microservice architecture for cloud applications grows in popularity, cloud services are becoming increasingly complex and more vulnerable to misconfiguration and software bugs.
Traditional approaches rely on expert input to diagnose and fix microservice anomalies, which lacks scalability in the face of the continuous integration and continuous deployment (CI/CD) paradigm. Microservice rollouts, containing new software installations, have complex interactions with the components of an application. Consequently, this added difficulty in attributing anomalous behavior to any specific installation or rollout results in potentially slower resolution times.
To address the gaps in current diagnostic methods, this paper introduces Praxium, a framework for anomaly detection and root cause inference. Praxium aids administrators in evaluating target metric performance in the context of dependency installation information provided by a software discovery tool, PraxiPaaS.
Praxium continuously monitors telemetry data to identify anomalies, then conducts root cause analysis via causal impact on recent software installations, in order to provide site reliability engineers (SRE) relevant information about an observed anomaly. In this paper, we demonstrate that Praxium is capable of effective anomaly detection and root cause inference, and we provide an analysis on effective anomaly detection hyperparameter tuning as needed in a practical setting. Across 75 total trials using four synthetic anomalies, anomaly detection consistently performs at $>0.97$ macro-F1. 
In addition, we show that causal impact analysis reliably infers the correct root cause of anomalies, even as package installations occur at increasingly shorter intervals.


\end{abstract}


\section{Introduction}
\label{sec:intro}
Modern cloud-native applications increasingly adopt microservice architectures for better maintenance and scalability.
In contrast to a traditional monolithic design, a microservice-based application is composed of multiple services running on different pods in a Platform as a Service (PaaS) environment.
These services communicate with each other using remote procedure calls (RPC).
With continuous integration and continuous deployment (CI/CD) practices, microservices provide applications with high flexibility of modification and scalability.

However, this flexibility complicates debugging and root cause tracing by site reliability engineers (SRE).
Performance anomalies can have varying root causes, such as improper configurations in a new version rollout, bugs in software dependencies, and faulty nodes, etc.~\cite{sage,Astraea}.
In order to explore the causal relationships, SREs have to investigate logs and metrics with very high dimensionality.
While new Agile practices and machine learning-based automation approaches gain traction, such manual diagnostics are no longer practical due to a lack of scalability.


Traditional root cause analysis in microservice systems often relies on service dependency graphs and alerts triggered by Service Level Objectives (SLOs) breaches. Tools like Jaeger~\cite{Jaeger} instrument services to generate spans, capturing task execution across components. Building on this, newer systems~\cite{tritium} incorporate machine learning to detect anomalies and infer root causes, using telemetry data from monitoring platforms such as Prometheus~\cite{prometheus} and Jaeger. Recent diagnostic methods further apply deep learning techniques, including variational autoencoders (VAEs), to improve the accuracy of anomaly detection and root cause inference~\cite{sage, RCD,cham2025}.

However, current approaches lack software-level granularity in root cause inference. Package installations in languages such as Python or JavaScript frequently introduce vulnerabilities, incompatibilities, or broken dependencies that can cause SLO violations in cloud environments. In order to detect the packages that may be at fault for a specific anomaly, administrators must comb through potentially hundreds of installation logs manually.

In order to address these issues, we introduce Praxium, a framework to employ microservice software installation information with an unsupervised learning anomaly detection approach to efficiently localize and diagnose the root cause of anomalies.

Praxium consists of three components.
First, a software discovery tool records the changes in dependencies during each microservice rollout with timestamps.
Second, Praxium continuously monitors key metric deviations from the normal behaviors modeled by a variational autoencoder. 
Third, once an anomaly is detected, Praxium examines the scenario by using a Bayesian model to generate counterfactual metric data of each affected service in the dependency graph. By comparing these counterfactuals to our observed telemetry, a prediction can be made on which software installation rollout contributed to the anomaly.


We make the following specific contributions in the design and implementation of Praxium:
\begin{itemize}
  \item We design a service metrics-based anomaly detection system using a VAE in the context of microservice rollouts.
  \item We integrate an ML-based system for logging and reporting service components and software changes after service rollouts, enhancing root cause analysis in microservices.
  \item We implement the comprehensive framework for anomaly root cause analysis and service dependency change logging in microservices using the New England Research Cloud (NERC) and OpenShift AI.
\end{itemize}

To the best of our knowledge, Praxium is the first work to connect dependency changes (container image changes) during microservice rollouts to anomaly detection and root cause analysis.
These dependency change logs localize the search space, effectively minimizing the exploration overhead.

This paper is structured as follows.
Sec.~\ref{sec:related} introduces the related work and the background motivating Praxium.
Sec.~\ref{sec:praxium} presents the system design of Praxium and explains the interactions between components.
Sec.~\ref{sec:eval} evaluates Praxium in terms of its anomaly detection and root cause analysis performance.
Finally, Sec.~\ref{sec:conclusion} concludes this work by discussing our system design, the evaluation, results, and further work.

\section{Related Work}
\label{sec:related}
Existing literature has extensively explored the topics of anomaly detection and root cause analysis. 
In this section, we summarize the monitoring tools helping SREs diagnose anomalous activity, the machine learning based anomaly detection and root cause analysis, and software bill of materials (SBOM) tools used to log software changes in a microservice rollout. 

\subsection{Causality Graphs}
A causality graph of microservices shows the upstream and downstream services that influence an anomaly detected by rule or machine learning based alarms set on performance issues.
This added information helps SRE to localize the root cause of anomalies~\cite{tritium}.

Previous works leverage tracing systems to construct a causality graph of microservices.
Tracing systems~\cite{Jaeger, Dapper, Zipkin, GMTA} monitor individual service calls, such as RPCs, and log performance metrics (i.e. running time) at each microservice with various granularities, known as \textit{spans}.
Thus, one can build a causality graph for each request with microservices as nodes and RPCs as edges to monitor anomaly propagation in the service~\cite{tritium}.
In contrast, recent works instead use \textit{Causal Bayesian Networks} (CBN) with performance metrics as nodes and metric causality, based on an RPC graph, as edges~\cite{sage}.


\subsection{Anomaly Detection}
Recent anomaly detection methods in microservice environments have increasingly applied machine learning techniques to identify deviations from normal execution patterns.
A common method is to model and generate normal tracing data with a VAE.
For example, TraceAnomaly~\cite{TraceAnomaly} encodes response times and invocation paths into a trace vector and employs a VAE to capture normal patterns, flagging anomalies based on low reconstruction probability for different paths. 
Similarly, TraceModel~\cite{TraceModel} utilizes causality graphs but restricts its analysis to the longest-running call path within each request's causality graph. Additionally, it constructs VAEs for different request categories to account for variations in the distribution of normal response times.


For HPC workloads, there are analogous anomaly detection tools measuring end-to-end performance, demonstrating that the approach generalizes across very different software stacks. An example is Prodigy~\cite{Prodigy}, which is an end-to-end, unsupervised VAE based framework that integrates performance metrics.

In this work, we utilize and expand upon Prodigy for microservice anomaly detection and root cause analysis, leveraging window-based metrics. This approach captures normal behavior in microservice metrics with fidelity, and adds robustness to anomaly detection when compared with traditional statistical methods used in similar frameworks ~\cite{tritium}.

\subsection{Machine Learning Based Root Cause Analysis}
ML-based approaches reduce manual diagnosis by learning normal behavior and simulating counterfactual scenarios to localize the faulty microservices in causality graphs.
After identifying an SLO violation, Tritium~\cite{tritium} localizes the area of effect in a causality graph based on metrics.
The framework then identifies faulty services using counterfactual analysis to simulate microservice rollouts and rollbacks. 
Similarly, Sage~\cite{sage} employs a CBN to model and represent the service causality graph of all microservice metrics.
The CBN consists of individual Conditional VAEs that each generate counterfactual metric data, enabling efficient incremental training. 
In contrast, Root Cause Discovery (RCD)~\cite{RCD} adopts a hierarchical causal discovery approach that uses conditional independence tests in subgraphs to efficiently narrow down candidate metrics.

While prior work focuses on microservice rollouts and trace data, Praxium takes a finer-grained approach by instead targeting software installation events within microservice rollouts. Whereas broader rollout operations may include changes in deployment patterns, scaling behavior, or new service versions, Praxium isolates potentially problematic software dependency-level changes to better localize the source of anomalies.

\subsection{Software Bill of Materials}
Logging software changes after each service rollout helps to localize the affected services when an anomaly is detected.
Traditionally, software discovery relies on the package management system. For example, we can query an installation database for package installation~\cite{osv-scanner, snyk}.
Another strategy is to rely on a system architect or SRE to pre-define a set of rules to identify software installation~\cite{OWASP}.
For example, since ``libcurl-750590ea.so.4.7.0" and ``libcurl-fiona4ac9f96f.so.4.8.0" are unique to ``fiona==1.9.4" and ``fiona==1.9.5", identifying if either file exists can help infer the installation of the corresponding package~\cite{PraxiPaaS}.
However, such approaches are fragile to maintain and do not apply when the installation does not use a package management system.
An ML-based approach for software change discovery at service rollout would train quickly, and be able to infer different versions of the software~\cite{DeltaSherlock, Praxi, PraxiPaaS}.
In PraxiPaaS~\cite{PraxiPaaS}, researchers compare pathname changes during package installation to identify packages and their versions.
The system was developed for a platform-as-a-service (PaaS) environment, thus it reads file changes from container image layers, which we use as a database to localize area of affect in the service.

\subsection{Vision for Praxium}


While prior work has focused on microservice rollouts, deployment patterns, and trace-based diagnostics, Praxium’s goal is to enable fine-grained root cause analysis by linking software installation events with telemetry-based anomaly detection. Unlike systems that rely solely on service-level traces or high-level rollout data, Praxium introduces three key novel aspects: (1) it connects software discovery mechanisms with telemetry analysis to support debugging of installation-induced anomalies; (2) it achieves higher diagnostic precision by examining package-level installation logs rather than coarse-grained rollout data; and (3) it is implemented and evaluated on a real-world Kubernetes-based research cloud, NERC, demonstrating practical feasibility through proof-of-concept experiments.

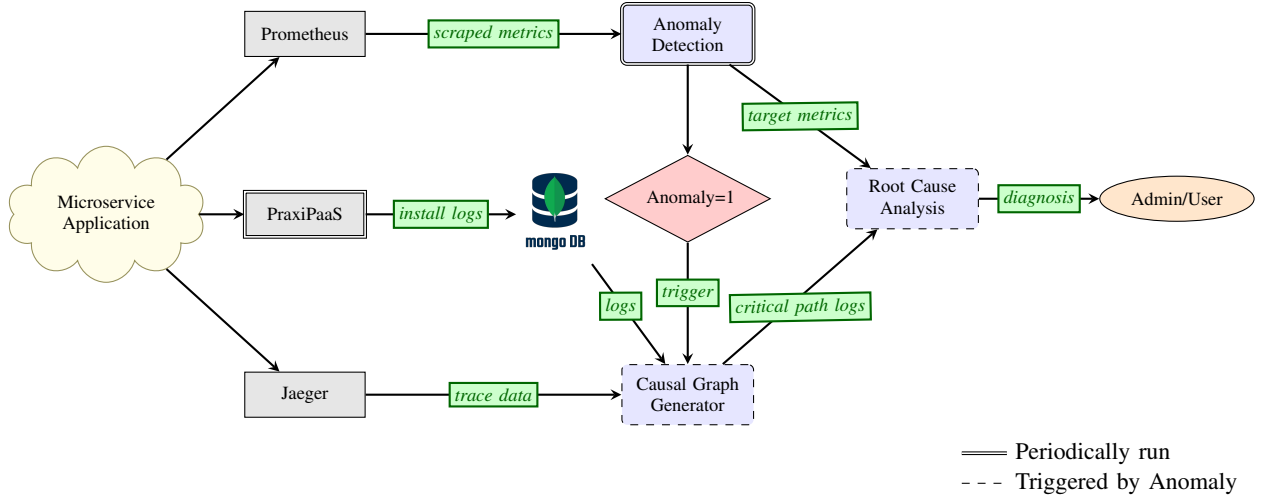
\begin{figure*}[!h]

    \centering
    \begin{tikzpicture}[
  node distance = 0.8cm and 1.8cm, 
  every node/.style = {font=\small}
]

\tikzstyle{tool} = [rectangle, minimum width=1.6cm, minimum height=0.6cm, text centered, draw=black, fill=gray!20, font=\scriptsize]

\tikzstyle{component} = [rectangle, rounded corners=2pt, minimum width=1.75cm, text width=1.4cm, minimum height=0.8cm, text centered, draw=black, fill=blue!10, font=\scriptsize]

\tikzstyle{trigger} = [diamond, aspect=1.9, draw=black, fill=red!20, minimum width=0.4cm, text centered, text width=1.1cm, font=\scriptsize]

\tikzstyle{user} = [ellipse, minimum width=0.4cm, minimum height=0.6cm, text centered, draw=black, fill=orange!20, font=\scriptsize]

\node (cloud) [cloudnode] {Microservice Application};

\node (prometheus) [tool, right=0.6cm of cloud, yshift= 2.4cm] {Prometheus};
\node (praxi)      [tool, double, right=0.6cm of cloud] {PraxiPaaS};
\node (jaeger)     [tool, right=0.6cm of cloud, yshift=-2.4cm] {Jaeger};

\node (anomaly) [component, double, right=3.4cm of prometheus]
      {Anomaly Detection};

\node (mongo)   [anchor=center, right=2.0cm of praxi]
      {\includegraphics[width=0.8cm]{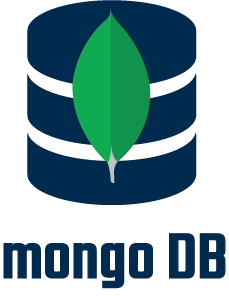}};

\node (graphgen) [component, dashed, right=3.4cm of jaeger] {Causal Graph Generator};

\node (trigger)  [trigger, below=1.2cm of anomaly] {Anomaly=1};
\node (ci)       [component, dashed, right=1.0cm of trigger]
      {Root Cause Analysis};

\node (user) [user, right=1.6cm of ci] {Admin/User};


\node[anchor=north east] at (current bounding box.south east) {
    \begin{tikzpicture}[scale=1]
      \draw[-, double] (0,0) -- (0.6,0) node[right] {Periodically run};
      \draw[-, dashed] (0,-0.4) -- (0.6,-0.4) node[right] {Triggered by Anomaly};
    \end{tikzpicture}
  };

\draw[arrow] (cloud) -- (prometheus);
\draw[arrow] (cloud) -- (praxi);
\draw[arrow] (cloud) -- (jaeger);

\draw[arrow] (prometheus) -- node[act,sloped]{scraped metrics} (anomaly);
\draw[arrow] (praxi)      -- node[act,sloped]{install logs}    (mongo);
\draw[arrow] (jaeger)     -- node[act,sloped]{trace data}          (graphgen);

\draw[arrow] (mongo) -- node[act, rotate=54, yshift=3pt, xshift=-3]{logs} (graphgen);

\draw[arrow] (anomaly) -- node[act, rotate=37]{target metrics} (ci);
\draw[arrow] (anomaly) -- (trigger);

\draw[arrow] (trigger.south) -- node[act,rotate=90,yshift=4]{trigger}(graphgen);

\draw[arrow] (graphgen) -- node[act, rotate=-40, yshift=-4]{critical path logs} (ci);

\draw[arrow] (ci) -- node[act]{diagnosis} (user);


\end{tikzpicture}
    \caption{An overview of Praxium. The system monitors a microservice cluster, collecting telemetry, trace, and installation data via Prometheus, Jaeger, and PraxiPaaS. Install logs are stored in MongoDB, while telemetry is streamed through anomaly detection. Upon detection of an anomaly, trace data populates the causal graph generator, and critical path logs along with target pod metrics pass to root cause analysis. From there, Praxium diagnoses the anomaly to a single installation.}
    \label{fig:prax_vision}
    
\end{figure*}

\section{Praxium}
\label{sec:praxium}

Praxium aims to serve as a unified tool capable of simultaneously tracking anomalies and installation logs, as well as correlating these anomalies with software installations when applicable. 
In doing so, this framework provides users and administrators with valuable diagnostic information during microservice rollouts.
Praxium is comprised of three systems that perform software dependency change logging, anomaly detection, and root cause analysis. 
Logging and anomaly detection are based on periodic intervals determined by an administrator. 
The anomaly detection collaborates with a trigger system that collects information to alert administrators about the positive identification of an anomaly in a target SLO. 
When anomaly detection triggers an alert, root cause inference examines all system information, including recent telemetry, recent installation logs, and recent microservice rollouts, in conjunction with causal graphs that provide service-level dependencies within the application. Figure \ref{fig:prax_vision} displays a high-level view of Praxium. Furthermore, Algorithm \ref{praxalg} details the system logic in pseudocode.

\begin{algorithm}[h]
\label{praxalg}
\caption{Praxium: Anomaly Detection and Root Cause Inference}
\KwIn{Telemetry $M(t)$, Traces $J(t)$, Install logs $I(t)$}
\KwOut{Anomaly report with causal installation (if found)}

\SetKwFunction{FProdigy}{Prodigy}
\SetKwFunction{FImpact}{CausalImpact}

Initialize $A[pod] \gets 0$ for all pods\;

\ForEach{sliding window $W_t$ over $M(t)$}{
    $predictions \gets$ \FProdigy{$W_t$}\tcp*[r]{Binary anomaly for each pod}

    \ForEach{pod $p$}{
        \eIf{$predictions[p] = 1$}{
            $A[p] \gets A[p] + 1$\;
        }{
            $A[p] \gets 0$\tcp*[r]{Reset if no anomaly}
        }

        \If{$A[p] \geq \tau$}{
            $t^* \gets$ current time\;
            $G \gets$ causal graph from $J(t^*)$\;
            $C \gets$ critical path in $G$ for pod $p$\;
            $T_C \gets$ install timestamps in $I(t^*)$ on $C$\;

            \ForEach{$t_i \in T_C$}{
                effect, $pval \gets$ \FImpact{pre/post-metrics around $t_i$}\;
            }

            \Return{timestamp $t_i$ with highest effect and $pval < 0.01$}\;
        }
    }
}

\end{algorithm}

\subsection{Software Dependency Logging System}
In a microservice-based application, the components can dynamically change due to CI/CD practices. For example, in a research computing cluster, CI/CD pipelines could be routinely deploying updated versions of data processing microservices, such as the software dependencies of model trainers or feature extractors. This could result in dynamic and frequent changes to the system’s runtime composition without downtime.

To minimize the overhead of running SBOM generation tools, we design a periodically triggered logging system.
For a fixed time interval, we retrieve the new or modified services, discover all corresponding new software, compare to the last inspection, and save that information as dictionaries in persistent storage.

Our software logging system utilizes the insights of the open-source PraxiPaaS~\cite{PraxiPaaS}. PraxiPaaS offers automated and scalable SBOM inspection, via a supervised machine learning inference pipeline to predict software package installations. 
To accomplish SBOM generation, we adapt PraxiPaaS into a minimal setup, and configure the Kubeflow pipeline to periodically run on a PaaS.
However, other alternative tools can be used.


\begin{figure}[t]
    \centering
    \includegraphics[width=1.0\linewidth]{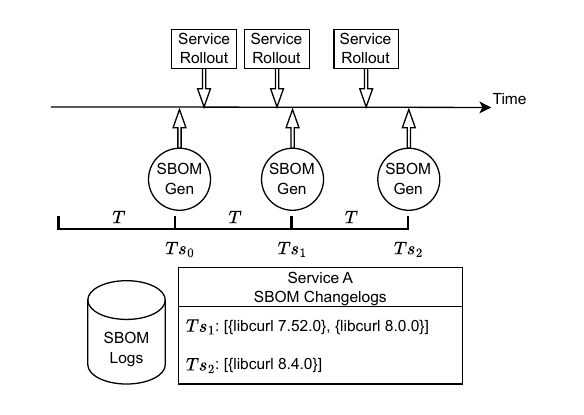}
    \caption{An example of the software dependency logging system. A background process periodically scans for software changes that occurred within the past $T$ time units. For each service, it creates a SBOM log history in a persistent database.
    In the figure, there were $3$ rollouts for Service $A$ and each of them updates the $libcurl$ lib.
    The first generation found no changes to Service $A$ in the past $T$ unit time, so no entry is created in the database.
    The second one can see two rollouts, so it create an entry with the timestamp ($Ts_{0}$) as the key where the value is a list of sets including software changes in deployment order. 
    Note that the first set includes all software of the service but the second set only has the changed ones.
    The final run can see the last rollout and creates a corresponding entry in the database.
    }
    \label{fig:SBOM-Logging}
\end{figure}

Praxium finds redeployment timestamps to minimize the frequency of software discovery operations. By conducting daily or weekly scans of PraxiPaaS, the system efficiently captures timestamped installation events. As shown in Fig.~\ref{fig:SBOM-Logging}, the software bill of materials generation can be called with an interval $T$, but the timestamps of service and pod deployments (e.g., $T_{s_{1}}$, $T_{s_{2}}$) can be captured by the system. This strategy reduces data collection overhead and enhances the tractability of root cause analysis in persistent PaaS environments.



\subsection{Anomaly Detection and Triggering System}
The triggering system keeps track of the latency of each application in the microservice cluster and identifies anomalous behaviors, e.g., SLO violations. First, we collect telemetry via a PaaS monitoring system such as Prometheus. 
We adjust Prodigy~\cite{Prodigy} for the microservice anomaly detection task, which establishes a VAE-based anomaly detection framework for HPC jobs.
In order to adapt Prodigy for the microservice architecture, we modify the model to output a binary prediction on the occurrence of a metrics anomaly within that pod over the past interval, for every pod within a microservice application. 


The anomaly detection is based on a series of sliding windows of microservice metrics.
In the offline training phase, we train a VAE model using observed healthy metrics of each pod in the current rollout. We use metrics available via prometheus node exporter like container\_memory\_usage\_bytes, container\_cpu\_usage\_seconds\_total, etc.
First, we generate a series of sliding windows with window stride as $S$ and window size as $W$. 
After this, a VAE model is trained for all pods simultaneously, using those sliding windows in chronological order. During training, the VAE reconstructs the extracted features from the metric data and then sets a threshold for the reconstruction error.
In the online phase, we treat the observed metrics as a stream processing scenario.
Praxium creates a sliding window view of the performance metric and uses the pretrained VAE models to reconstruct the metric data.  If the reconstruction error passes the set threshold determined in training, the system detects an anomaly in that window.
When there are $\tau$ number of such consecutive detected anomalous windows, the anomaly detection unit reports the anomaly to further components, including the time of the event and the affected pod/service. 

Training individual VAEs for each pod can be challenging when leveraging matrix multiplication optimized hardware architectures, such as GPUs.
Considerations include different formatting and availability of metrics between pods, complicating the process of training on a single GPU.
For better training and inference parallelism, we build a single VAE model and concatenate metrics of all pods in each window as a single data frame with a notation for individual pods. 
Compared to using individual VAE for each pod, our approach can minimize the parameter migration overhead during forward and backward propagation.

\subsection{Root Cause Analysis and Causal Graph}
Before attempting to correlate recent installations to an observed anomaly, Praxium first builds a causal graph from Jaeger spans of the current application. 
This causal graph essentially models upstream-downstream dependencies between sub-services. 
In the case of an anomalous service, Praxium only considers installations made in the critical path of that service, where the critical path is defined as all upstream and downstream services. This has two effects: this expands the candidate installations that could be the root cause, while simultaneously limiting the number of installations to consider, as Praxium can then discard all installations outside of the critical path. Once Praxium finalizes this subset of installations, it performs root cause analysis. 

To identify the deviation of latency from norm, we utilize CausalImpact~\cite{causal-impact}, a Bayesian counterfactual generator algorithm and hypothesis tester, that functions as a root cause inference mechanism. At the instance of a detected anomaly, the metrics collected in the previous window are passed to this model, as well as the timestamps of recent microservice rollouts. Here, the model attempts to correlate these timestamps with problematic telemetry by generating counterfactual data starting at each timestamp and performing posterior inference on the artificial data and the observed anomalous data. By observing correlations between different metrics found in the healthy data prior to the anomaly, the model produces confidence intervals to show how significant the impact of each timestamp is on telemetry changes. To correlate the anomaly with a single timestamp, Praxium first collects all timestamps with a p-value below 0.01, then returns the timestamp of this subset that has the highest average effect. In this way, the system finds the most likely cause of the anomaly from the set of candidates.



\section{Evaluations}
\label{sec:eval}

Evaluation consists of multiple controlled trials that test the effectiveness of our anomaly detection and causal inference framework under various configurations and system conditions. 
The primary objective of evaluation is to measure Praxium's ability to detect anomalies, correctly identify the affected pod, and determine whether causal inference improves root cause analysis. 
In addition, we analyze various hyperparameter configurations to offer insights into sensitivity and consistency for administrators utilizing this tool. Subsection~\ref{subsec:method} provides an overview of our experimental design, including information about the system conditions and synthetic anomalies. Subsections~\ref{subsec:detection} through~\ref{subsec:graph} each detail one of three evaluations, and finally, subsection~\ref{subsec:disc} reviews and discusses the results of these experiments in the context of future work.

\subsection{Experimental Methodology}
\label{subsec:method}

We deploy a Kubernetes-based microservice architecture, DeathStarbench - Social Network \cite{deathstarbench}, on OpenShift, and collect system telemetry data under healthy and anomalous conditions. 
The Social Network application models a one-way follower system for content broadcasting, including load balancing, persistent storage, and user interfaces. 
Within NERC, Prometheus logs metrics every 30 seconds, thus we choose this as the minimum window stride value.
During evaluations, anomalies are injected via a package installation.
Our software discovery and installation log generation system captures these changes  and stores them as logs to be used later in diagnosing the root cause.
We introduce programs capable of spiking resource utilization patterns to the PyYAML~\cite{pyyaml} package source code, after which we can redeploy a specific service from an altered deployment configuration file containing that new version of PyYAML. With varying anomalous deployments, we evaluate our system when an anomalous package is installed via an update to the pods' containers.
Table~\ref{tab:anomalies} contains information regarding the four synthetic anomalies used in evaluation. 

In addition, we initiate a background load throughout the Social Network service via wrk2~\cite{wrk2} while experimenting with different anomalies. Wrk2 generates a throughput load for the Social Network service by instantiating different application functions. For example, we choose the \textit{ComposePost} function, which creates a post, utilizing several different services in the process.
Thus, the base telemetry is non-zero and non-constant, providing a better representation of a real-world application.

\begin{table}[t]
    \caption{Synthetic Anomalies Injected During Experiments}
    \renewcommand{\arraystretch}{1.2}
    \label{tab:anomalies}
    \centering
    \begin{tabularx}{\columnwidth}{l X}
        \toprule
        \textbf{Injected Anomaly} & \textbf{Injection Method and Target Component} \\
        \midrule
        CPU Spike & A script triggers CPU-intensive computations inside the target pod to simulate high CPU usage. \\
        Disk Saturation & Temporary files are written to exhaust available disk space on the pod's storage volume. \\
        Memory Leak & The application is instrumented to continuously allocate memory without releasing it. \\
        Network Latency & Repeated outbound HTTP requests are issued from the pod to flood its network interface. \\
        \bottomrule
    \end{tabularx}
\end{table}

\subsection{Anomaly Detection: Hyperparameter Tradeoffs}
\label{subsec:detection}

\begin{figure}[b]
    \centering
    \includegraphics[width=1.0\linewidth]{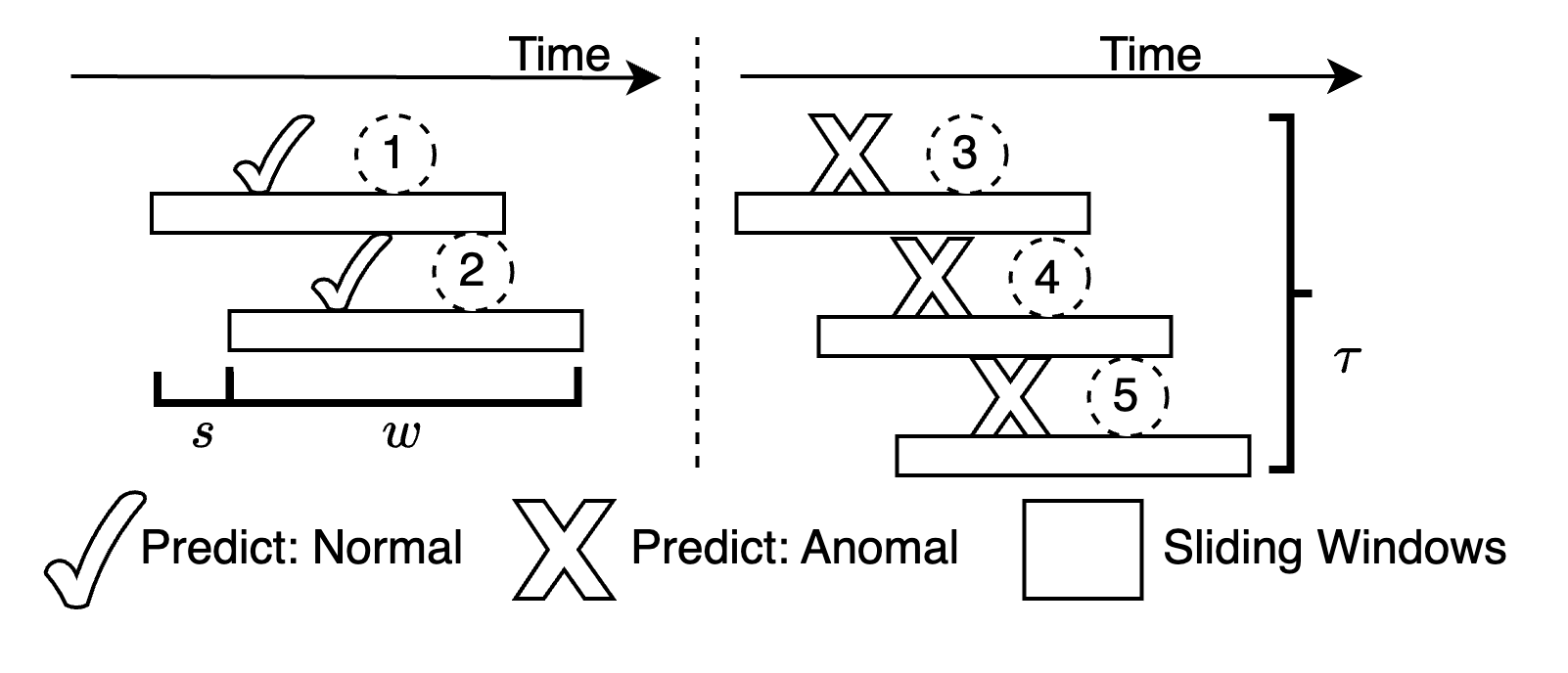}
    \caption{An example of the hyperparameters of experiment 1. Stride $s$ controls the distance of time shifted between each window. Window size $w$ is the duration of each window. Threshold $T$ is the number of consecutive anomalous windows necessary to alarm the system about an anomaly. }
    \label{fig:exp1_fig}
\end{figure}

\begin{table*}[ht]
    \caption{Best Window Duration (W) and Stride (S) Configuration with Performance Metrics for Each Anomaly Type}
    \label{tab:best_anomaly_results}
    \hspace*{\fill}
    \begin{minipage}{\textwidth}
    \centering
    \begin{tabular}{l|rr|rrrr|rrrr}
    \toprule
    \multirow{2}{*}{\textbf{Anomaly}} 
        & \multicolumn{2}{c|}{\textbf{Best W, S (sec)}} 
        & \multicolumn{4}{c|}{\textbf{No Thresholding (T=1)}} 
        & \multicolumn{4}{c}{\textbf{With Thresholding (T=2)}} \\
        & W & S  
        & F1 & Precision & Recall & Accuracy 
        & F1 & Precision & Recall & Accuracy \\
    \midrule
    CPU  & 600 & 300  & 0.985 & 0.972 & 0.998 & 0.996 & 1.0 & 1.0 & 1.0 & 1.0 \\
    RAM  & 600  & 300  & 0.961 & 0.932 & 0.994 & 0.989 & 1.0 & 1.0 & 1.0 & 1.0 \\
    DISK & 600 & 30  & 0.951 & 0.941 & 0.960 & 0.978 & 0.976 & 0.987 & 0.961 & 0.990 \\
    HTTP & 300  & 60  & 0.958 & 0.932 & 0.988 & 0.936 & 0.995 & 0.999 & 0.991 & 0.999 \\
    \bottomrule
    \end{tabular}
    \end{minipage}
    \hspace*{\fill}
\end{table*}

The first experiment evaluates how well the system detects and localizes anomalies under different systems conditions and hyperparameters. 
Three key hyperparameters influence the anomaly detection system:

\begin{enumerate}
    \item \textbf{Window Duration ($W$)}: Length of the telemetry interval that is considered when checking for an anomaly. Larger windows smooth out short-term fluctuations but may delay detection. We try 600 seconds, 450 seconds, 300 seconds.
    \item \textbf{Window Stride ($S$)}: Length that a window advances after each computation. When $S \le W$, there are overlapping windows in the anomaly detection, which improves the granularity of anomaly detection but introduces redundant computations. When $S \geq W$, we have non-overlapping windows, which reduce computation overhead, but might overlook anomalies incurred in between windows. We try 300 seconds, 60 seconds, and 30 seconds.
    \item \textbf{Detection Thresholding ($T$)}: Number of consecutive windows that must be flagged as anomalous before an anomaly is reported. A high threshold reduces false positives, but may also increase the detection delay and hide intermittent anomalies. We try with and without thresholding (i.e. $T=1$ and $T=2$).
\end{enumerate}
Figure \ref{fig:exp1_fig} visualizes these hyperparameters as an example of how Praxium will perform window-based anomaly detection.

We evaluate the system across different values of $W$, $S$, and $T$ to assess their impact on performance.
All four anomalies are used as systems conditions in varying degrees of utilization. For example, in each run of the CPU utilization anomaly experiments, we randomly choose a utilization percentage between $0$ to $100$. In this way, we are able to simulate varying degrees of system disruption.

We train our Prodigy-based model on healthy telemetry data that span $18$ hours. From these $18$ hours, we create windows based on the hyperparameters. In inference, each anomaly is instantiated 25 times, each time in a randomly selected pod.
We evaluate detection performance using F1-score, precision, recall, and accuracy. We report weighted averages of each performance metric based on the percentage of anomalous to healthy windows in our test set (i.e. macro-F1). 

In Table \ref{tab:best_anomaly_results}, for readability, we only report the best hyperparameter configuration for each anomaly. We find these optimal configurations by performing grid-search on all hyperparameters, for each anomaly. 
The best configuration is the combination of $W$ and $S$ that provides the highest F1 after thresholding. In the case of a tie, then the combination of $W$ and $S$ that provides the highest F1 before thresholding is used. Overall, our anomaly detection method successfully detects anomalies, evidenced by the F1 scores and accuracy metrics that reach $>0.97$ consistently after thresholding. In the case of the CPU and RAM anomalies, all performance metrics reach 1.0 (when rounded to 3 significant figures). Thresholding allows false positive filtering, as shown by the improvements in precision. For example, in the CPU anomaly case, precision before thresholding is $0.972$, and afterwards, is 1.0. This demonstrates that anomalies were being predicted due to noise in single windows, and so thresholding fixed this issue. Another important observation is for administrators choice of hyperparameters. While thresholding seems always beneficial, window duration and stride vary in effectivity between anomalies. 
However,
the model with 600 second (10 minute) window duration and 300 second (5 minute) stride consistently provides high performance in the case of CPU and RAM anomalies, and is only marginally worse than other hyperparameter configurations for all four anomalies. There is another tradeoff in the overhead of the model, as smaller strides lead to more frequent anomaly detection runs. For this reason, a 600-second window duration with a 300-second stride is recommended to balance detection granularity and system efficiency.

\subsection{CausalImpact Analysis}
\label{subsec:impact}

In this experiment, we examine whether integrating CausalImpact improves the system’s ability to determine whether a package installation or microservice rollout was the root cause of an anomaly. 
The naive approach would report the most recent installation to the time of the anomaly. 
However, as window size and slide interval of anomaly detection increases, this approach provides less and less reliable information. 
If an anomalous window contains multiple installations, such as what might occur within an Agile-style development environment, it may be the case that the most recent installation is not at fault. 

As described in the system design, Praxium only begins root cause analysis when it detects an anomaly and an installation event has occurred within a recent time frame (otherwise, we do not predict any root cause).
Once these conditions are met, we evaluate whether our system can accurately pinpoint which of several recent installations best correlates to the observed anomaly, via CausalImpact. 
For each trial, we perform 5 separate installations into a pod, with one of these installations also injecting an anomaly. 
Then, once all installations have occurred, CausalImpact is used to predict which, if any, of these installation timestamps best corresponds to the anomalous telemetry.

\begin{figure}[t]
    \centering
    \includegraphics[width=1.0\linewidth]{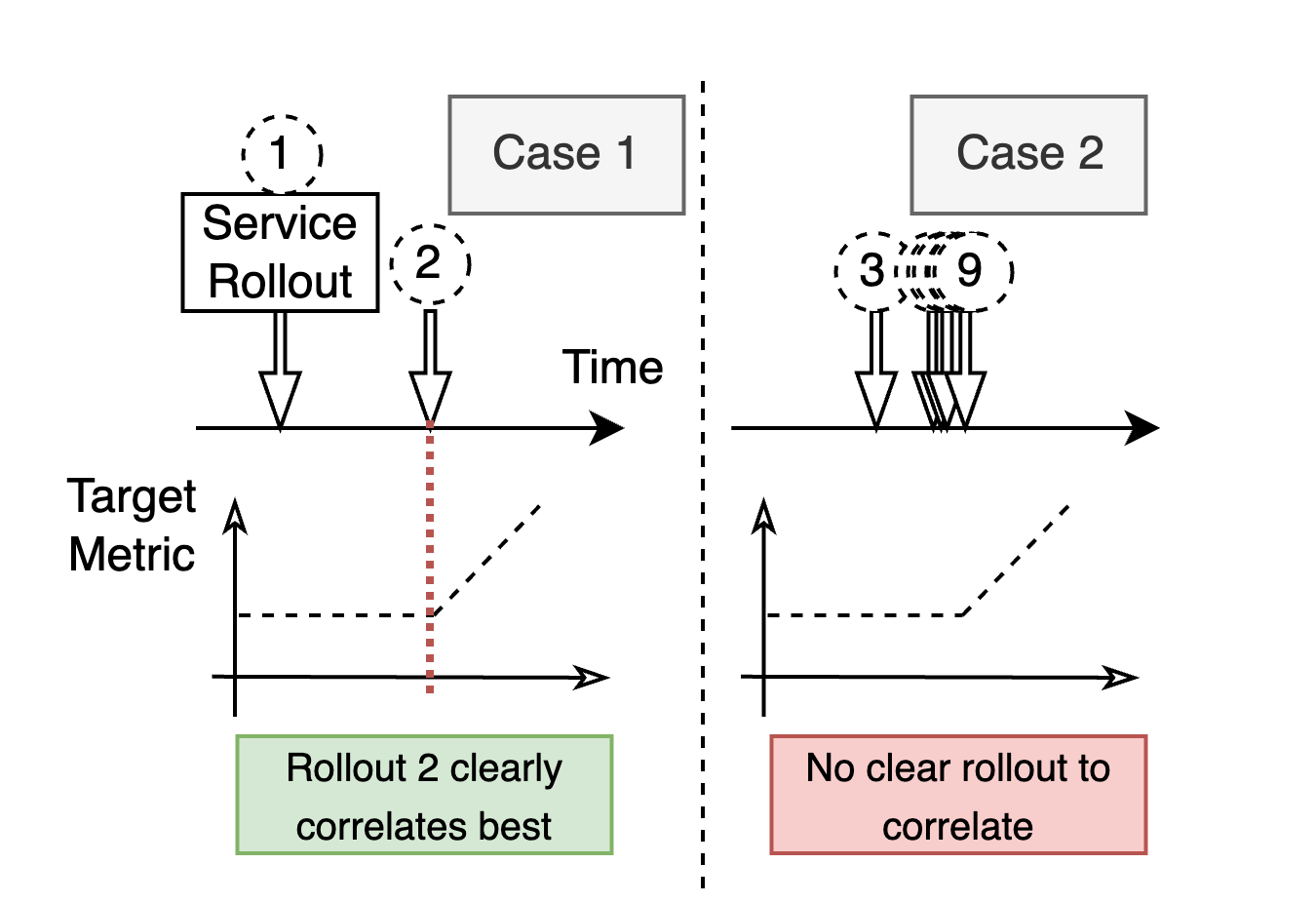}
    \caption{An example of experiment 2, in two cases. First, the case 1 is trivial: there is an easily identified rollout that correlates to the shift in target metric. Then, in case 2, the quick succession of rollouts crowds around the shift in metric, obfuscating the best-correlated rollout.}
    \label{fig:exp2_fig}
\end{figure}

This experiment varies the interval time between deployments. 
The goal of this is to examine how well CausalImpact, and consequently Praxium, handles deployments that are close together. 
To do this, we choose intervals of 10 minutes, 5 minutes, and 2 minutes. For each, we conduct $3$ trials. In practice, these intervals correspond to the stride length of the anomaly detection window: a longer stride increases the likelihood that multiple deployments fall within a single anomalous window. A trial is a success if the described algorithm for installation correlation reports the actual anomalous injection timestamp. Figure \ref{fig:exp2_fig} depicts an example of the experiment, as well as a visual explanation of the difficulty of inference when deployments are bunched together. Across all intervals, we observe that every trial is successful, such that across all 9 trials, causal inference identified the correct anomalous installation log. These results demonstrate that the algorithm effectively performs root cause analysis in a practical context. Furthermore, the ability to accurately identify root causes even with grouped installations supports the use of longer strides, which in turn reduces monitoring overhead.

\subsection{Causal Graph Analysis}
\label{subsec:graph}

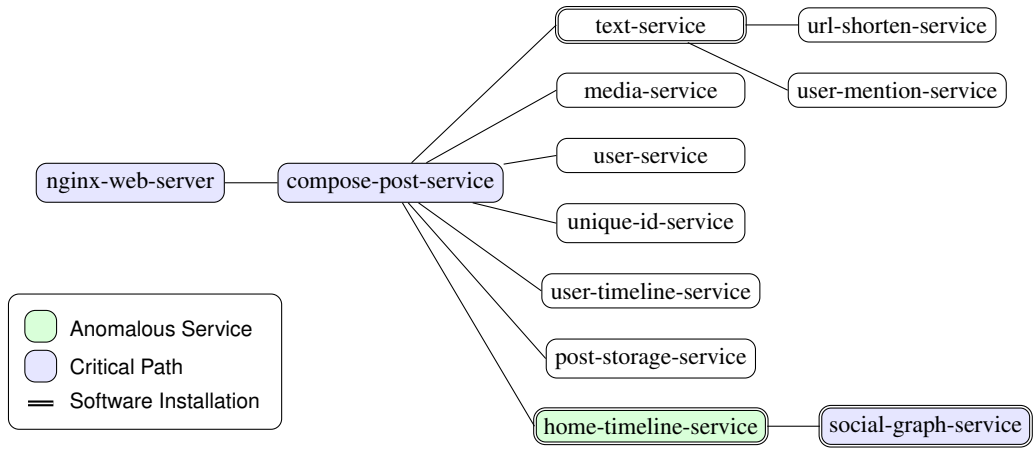
\begin{figure*}[!h]

    \centering










\begin{tikzpicture}[
  >={Stealth},
  node distance=0.4cm and 0.7cm,
  every node/.style={draw, rounded corners, font=\small, minimum width=2.5cm, align=center},
  every edge/.style={->, thick}
]

\node[fill=blue!10, draw] (nginx) {nginx-web-server};

\node[fill=blue!10, draw] (compose) [right=of nginx] {compose-post-service};

\node (text) [right=of compose, yshift=2.1cm, double] {text-service};
\node (media) [below=of text] {media-service};
\node (user) [below=of media] {user-service};
\node (uid) [below=of user] {unique-id-service};
\node (userTimeline) [below=of uid] {user-timeline-service};
\node (postStorage) [below=of userTimeline] {post-storage-service};
\node[fill=green!15, draw, double] (homeTimeline) [below=of postStorage] {home-timeline-service};

\node (url) [right=of text] {url-shorten-service};
\node (mention) [below=of url] {user-mention-service};
\node[fill=blue!10, draw, double] (socialGraph) [right=of homeTimeline] {social-graph-service};

\draw (nginx) -- (compose.west);
\draw (compose) -- (text.west);
\draw (compose) -- (media.west);
\draw (compose) -- (user.west);
\draw (compose) -- (uid.west);
\draw (compose) -- (userTimeline.west);
\draw (compose) -- (postStorage.west);
\draw (compose) -- (homeTimeline.west);
\draw (text) -- (url.west);
\draw (text) -- (mention.west);
\draw (homeTimeline) -- (socialGraph.west);



\node[anchor=north west, draw, fill=white, rounded corners, font=\footnotesize\sffamily, inner sep=6pt] at ([xshift=-7cm, yshift=2cm]homeTimeline.south west) {
  \begin{tabular}{@{}c@{\hspace{0.6em}}l@{}}
    \tikz \node[fill=green!15, draw, minimum height=0.8em, minimum width=1.2em] {}; & \raisebox{0.3em}{Anomalous Service} \\
    \tikz \node[fill=blue!10, draw, minimum height=0.8em, minimum width=1.2em] {}; & \raisebox{0.4em}{Critical Path} \\
    \tikz \draw[-, thick, double] (0,0) -- (1.2em,0); & \raisebox{-0.1em}{Software Installation} \\
  \end{tabular}
};

\end{tikzpicture}
    \caption{An example of the Social Network service dependency graph during the \textit{ComposePost} functionality. Here, an anomaly is injected into \textbf{home-timeline-service} while \textbf{home-timeline-service}, \textbf{social-graph-service}, and \textbf{text-service} are redeployed with new installations. Only installations along the critical path (in blue) are considered for causal inference.}
    \label{fig:service_dependency_example}
    
\end{figure*}

We extend the root cause analysis by incorporating causal graph dependencies. When an upstream service triggers an anomaly, it can lead to increased latency in a downstream service, potentially causing anomalies in pods that have not undergone recent installations. In such cases, Praxium should identify and report the installations of upstream services when anomalous latency is detected in the downstream service.
To evaluate this scenario, we simulate pod overloads along the critical path of a service, while also introducing installations to non-critical path services. We then use the causal graph to determine which installation logs should be considered: only those within the critical path are expected to be included for causal inference. Figure~\ref{fig:service_dependency_example} illustrates the structure of this scenario. The goal of this evaluation is to showcase Praxium’s practical applicability. Given that rollouts often involve multiple installations across various pods, the causal graph’s ability to accurately identify relevant installations is crucial for optimizing utility while minimizing overhead.

We observe 3 trials of this scenario, where for each trial, a pod is randomly chosen and different downstream pods are redeployed with installations. The causal graph used to identify these downstream pods are manually obtained, and then compared to the output of the causal graph generation during evaluation. Then, we perform causal impact on the subset of installation logs returned, and finally compare the returned culprit installation log with the ground truth. In each of the 3 trials, we are able to correctly build a causal graph from trace data, as well as identify the correct cause of the anomaly.

\subsection{Discussion and Future Work}
\label{subsec:disc}

\textbf{Anomaly definition and controlled stress scenarios}: Our evaluation focuses on unusual delays or utilization associated with a newly deployed container/pod, which may originate within that pod itself or within containers/pods in its upstream or downstream dependencies. To make this diagnosis setting measurable and repeatable, we use synthetic and controlled stress scenarios (CPU, memory, disk, and network) that inject load to emulate common resource pressure patterns and to demonstrate Praxium’s ability to localize telemetry changes correlated with recent installation events. However, these controlled injections do not capture the full diversity, timing, and concurrency of real production incidents, and we do not claim that performance on these scenarios fully generalizes to all real world failures. In practice, anomalies have broader definitions and may arise from configuration errors, scheduling or permission issues, node level faults, or external platform events; for example, misclassifications of nodes due to hardware availability or permission requirements can lead to denial of service. We view such cases as complementary anomaly types where similar time series anomaly detection and causal graph based root cause analysis may still be applicable, but they require additional modeling and validation beyond the scope of this evaluation.

\textbf{Interval in Observability and Praxium Windowing}: In our evaluated platform-as-a-service environment, Prometheus has a 30-second interval for scraping metrics. Deployments within a 30-second interval are unlikely to be distinguishable in metric data. An extension to our evaluation is to include finer granularity in metrics and deployment intervals. There will be a limit when the causal inference approach cannot correlate deployments in a window with anomalies. However, such a limit is dependent on the cloud system configuration. We introduce Praxium as a effective framework, and the specific implementation should be adapted to the user's use case.

\textbf{Effects of Tracing sampling and causal graph completeness}: In production systems that sample traces (for example, at 1\%), sampling may reduce the completeness of the causal graph. Praxium’s filtering step relies on this graph to narrow down likely root causes, so lower trace coverage can make the graph less informative and reduce Praxium’s ability to collect correct set  of candidate installations. That said, large scale systems that generate billions of traces per day still collect millions of traces even at 1\% sampling, which can reasonably represent the causal relationships between services and will be enough for Praxium to work.

\textbf{Large Scale Praxium Implementation}: Praxium is currently implemented as a Kubeflow pipeline that is triggered periodically.  Our evaluation application (DeathStarbench - Social Network) is also a small benchmark with a synthetic workload. We train a single VAE model for all our microservices(32 pods) taking under 10 minutes, We currently advise to use around 30-50 pods per VAE for similar results. One future extension of Praxium is to understand the practicality of Praxium in a large-scale microservice cluster (with thousands of pods) in the real world, including the running cost (computation and storage demand) and running time of the Praxium system.

\section{Conclusion}
\label{sec:conclusion}

In this work, we introduce Praxium, a framework for anomaly detection and root cause inference that leverages installation logs and telemetry monitoring to diagnose issues within a cloud environment. Through evaluation, we demonstrate that Praxium provides an effective tool in the detection and localization of anomalous telemetry in a real-time setting, while also generating valuable insights into the underlying cause via CausalImpact and causal graph analysis.

We identified key hyperparameters in our anomaly detection machine learning model, window size, slide interval, and false-positive threshold, that impact system performance and overhead. Administrators can adjust these parameters to compromise between sensitivity, speed, and resource usage. Optimal configurations balance the trade-off between early detection and reducing false positives, with our findings suggesting a window size of 10 minutes, a stride of 5 minutes, and false-positive filtering with a threshold of 2 yield the best results for anomaly detection in cloud environments similar to that of our evaluation system. Moreover, causal impact supports the use of longer stride, such as 5 minutes, due to the displayed efficacy in the case of multiple closely-deployed installations. Lastly, we examine a practical scenario in which the causal graph generation is able to provide more nuanced information, by intelligently expanding the possible installations to consider.

While Praxium demonstrates strong performance, future work could explore further refinement in causal inference modeling. For example, additional system information, such as package dependency information, or code-level package usage, could provide further granularity into inferring the cause of an anomaly.

Overall, Praxium provides a robust framework for cloud SREs to both detect anomalies and trace back their root causes, offering a significant improvement in daily administration over traditional anomaly detection systems that operate in isolation. As cloud environments continue to grow in complexity, systems such as Praxium can play a crucial role in enhancing the reliability and stability of cloud-based applications.

\section*{Acknowledgements}
This research is supported by Boston University Red Hat collaboratory.


\bibliographystyle{IEEEtran}
\bibliography{references}

\end{document}